**Counting the costs of COVID-19: why future treatment option values matter**

**Adrian Kent**

**Department of Applied Mathematics and Theoretical Physics**

**University of Cambridge, Cambridge CB3 0WA, United Kingdom.**

**Email: apak@damtp.cam.ac.uk**

**Abstract:** I critique a recent analysis (Miles, Stedman & Heald, 2020) of COVID-19 lockdown costs and benefits, focussing on the United Kingdom (UK). Miles et al. (2020) argue that the March-June UK lockdown was more costly than the benefit of lives saved, evaluated using the NICE threshold of £30000 for a quality-adjusted life year (QALY) and that the costs of a lockdown for 13 weeks from mid-June would be vastly greater than any plausible estimate of the benefits, even if easing produced a second infection wave causing over 7000 deaths weekly by mid-September.

I note here two key problems that significantly affect their estimates and cast doubt on their conclusions. Firstly, their calculations arbitrarily cut off after 13 weeks, without costing the epidemic end state. That is, they assume indifference between mid-September states of 13 or 7500 weekly deaths and corresponding infection rates. This seems indefensible unless one assumes that (a) there is little chance of any effective vaccine or improved medical or social interventions for the foreseeable future, (b) notwithstanding temporary lockdowns, COVID-19 will very likely propagate until herd immunity. Even under these assumptions it is very questionable. Secondly, they ignore the costs of serious illness, possible long-term lowering of life quality and expectancy for survivors. These are uncertain, but plausibly at least as large as the costs in deaths.

In summary, policy on tackling COVID-19 cannot be rationally made without estimating probabilities of future medical interventions and long-term illness costs. More work on modelling these uncertainties is urgently needed.

1. Introduction

Many countries, including the UK, have imposed temporary lockdowns in response to the COVID-19 pandemic. These have almost certainly saved lives, in the sense that fewer people have died from COVID-19 during and immediately after the lockdowns than would have done so in the absence of any government-mandated lockdown.   They may also have directly cost lives, in the sense that more people may have died or will die from some non-COVID causes than would otherwise have been the case.   They have also been extremely costly economically, and these costs translate into effective lives lost in various ways: for example, unemployment adversely affects quality of life and life expectancy, while lower government revenues and lower citizens' incomes mean fewer resources for public and private health care.    None of these costs and benefits are easy to estimate, nor is there any consensus on how to compare health and death costs with monetary costs.

In a recent article Miles et al. (2020) bravely take on these challenges.   As they say, without estimates and some framework for comparing human lives and well-being with monetary costs, policy would be made in a vacuum.    The approach they take is to consider the costs of COVID-19 deaths (or the benefits of lives saved by interventions) by estimating the quality-adjusted life years (QALYs) lost (or gained).   They then use one of the National Institute for Clinical Excellence (NICE) guideline (Paulden, 2017) figures, according to which a treatment expected to increase life expectancy by one QALY should cost no more than £30000.   Taking this as the value of a QALY allows them to compare the medical benefits of interventions with their financial costs.

Everything about this approach is debatable.    Some might argue that government decisions about dealing with pandemics or other disasters should not be guided by the same principles that are used in health care policy; that different values and criteria come into play.   Battling the pandemic has been widely compared to a war.   It would actually be very interesting to consider what QALY-based analyses would have said about the U.K.'s decisions to declare war in (for example) 1939 or 2003, but most people (then or now) probably would not think this is how the decisions should have been made.

These large questions are beyond my scope here.   For the sake of discussion, I will accept Miles et al.'s overall approach as one plausible way of getting insight into the policy questions.    As Miles et al. note, arguments can be made for using different values for a QALY, and their calculations can be applied with any value (though of course the conclusions depend on the value chosen). In fact, NICE themselves use a range of figures in different circumstances: their "end-of-life" guidance allows treatment expected to increase life expectancy by one QALY costing up to £50000, and they allow a threshold of up to £300000 per QALY for treating "very rare diseases" (Paulden, 2017). While COVID-19 is, alas, not a rare disease in the standard sense, one might interpret this larger threshold to reflect a societal wish for a particularly compassionate response in unusual circumstances -- such as, perhaps a global pandemic. On the other hand, the larger threshold might be considered affordable only for very rare diseases precisely because of their rarity.

I will in any case follow Miles et al. in using the NICE £30000 guideline as a primary reference point, while considering alternatives. My focus is on two narrower questions: does it make sense to follow Miles et al. in considering costs and benefits over a fixed period (13 weeks), without taking the future consequences into account? And is it justifiable to ignore the losses arising from COVID-19 illness that does not result in death? I argue that the answer to both questions is no, and hence that Miles et al.'s analyses are flawed and their conclusions unreliable.

## 2. End States Matter: Why a raging epidemic in mid-September is not cost-free

One does not have to dig deep into (Miles et al., 2020) to see a glaring problem with the analysis. Their table 4 sets out the deaths and corresponding (QALY-based) costs for a range of scenarios running over 13 weeks from 12.6.20. One is a continued full lockdown scenario, which they assume would reduce deaths and infections by a factor $F = 0.7$ per week, resulting in 13 deaths in week 13. The others are three "unlocking scenarios", in which restrictions are eased, resulting in the death rates multiplying by $F = 0.9, 1.0$ or $1.15$ each week, so that they slowly diminish, stay constant, or increase over time. The last of these gives a death rate of 7572 per week in week 13, close to the rate at the height of the UK epidemic in mid-April 2020. Miles et al. calculate the total number of deaths over weeks 1-13 in each of the four scenarios, use these to estimate the costs of the easing scenarios in QALYs, and compare the resulting figures with the economic cost of continued full lockdown for the 13 weeks.

The implication of this methodology is that we should be indifferent between the states we reach at the end of week 13 in the various scenarios, since Miles et al. assign no costs to any of them. We should, in particular, be indifferent as to whether we reach mid-September in a state of near suppression, with 13 weekly deaths (and falling), and a raging epidemic, with 7572 weekly deaths (and rising). We should care about the past costs – the deaths that have taken place – but ignore the future costs – the deaths that will predictably take place after mid-September as a consequence of the state at that time. I beg to differ, and I suspect most economists, epidemiologists, policymakers and health care professionals also would. The point is perhaps subtler than it seems at first sight, though, so it is worth considering various arguments:

*2.1 Good and bad reasons for caring about end states*

(1) *A raging epidemic second wave would make the sacrifices of the earlier lockdown pointless.*

> This sounds rhetorically forceful but is analytically unjustifiable. The sacrifices of the earlier lockdown are sunk costs. The lockdown may have been justified at the time because there was great uncertainty about the mortality and transmission rates of COVID-19, a real risk that an uncontrolled epidemic might lead to a complete collapse of the health care system, some hope that summer weather would reduce transmission enormously, or even conceivably because of some chance of a dramatic medical breakthrough early in the epidemic. We now know more, and in hindsight it is possible, and in fact in line with Miles et al.'s analyses, that the March-June lockdown was indeed unjustified, in the sense that the economic costs of the lockdown were greater than the benefits in QALYs saved**.** But in any case it is academic: we need to make future policies based on future costs and benefits, not on what they tell us about past policies.

*(2) For any given policy after mid-September, the death rates in subsequent weeks will be proportional to the death rates in week 13.*

> This would be true in a model with just two parameters, D, the initial death rate per week, and F, the constant factor by which this rate changes. In this model, the death rate in week N is $D\ F^N$, and the total deaths after N weeks are $DF \frac{(F^N - 1)}{(F - 1)}$. This is the model which Miles et al. use for a 13-week period, with $D = 1230$ and various values of $F$. Although obviously crude, it may be a reasonable way of setting out scenarios over this relatively short length of time. But death rates obviously cannot grow exponentially indefinitely, since the UK population is finite. After 39 weeks, with F = 1.15, this would give a cumulative total of 2,187,051 deaths, far more than the estimated total number of deaths if everyone in the UK were infected. If we want to discuss the future costs of a given mid-

September state, even only for the following six months, we clearly need a much more realistic epidemic model.

(3) *Mid-September is a particularly bad time to have a large and growing epidemic.*

The argument here is that, even if we must suffer the same cumulative death total eventually, it would be better to postpone the next wave beyond the winter if possible. This goes beyond my expertise but seems plausible. UK health care systems are reportedly very stretched over winter even in normal years. Influenza and other respiratory illnesses are more prevalent over the winter months. This takes up hospital space and resources. These illnesses, which have symptoms overlapping with those of COVID-19, also complicate self-diagnosis, and hence are likely to make testing slower and less complete, reducing the effectiveness of tracking, tracing, and quarantining.

(4) *The longer we can wait, the better chance we have of slowing, reducing and treating any second epidemic wave.*

This, I think, is the strongest argument. There is considerable uncertainty about when, if ever, a widely effective vaccine will be available, but there must be **some** chance (even if small) that it might be soon enough to mitigate a second wave in winter 2020-1.

There must also be some chance of better treatment options, even as soon as September 2020. The COVID-19 outbreak was declared a pandemic by the WHO in early March 2020. One might perhaps take this as a rough start date for global medical research focussing on COVID-19. It was already known by June 2020 that treating specific subsets of patients with dexamethasone appears to be effective in reducing mortality (Horby et al., 2020): it has been estimated that 4000-5000 COVID-19 deaths in the UK alone could have been avoided had this been known at the beginning of the epidemic (Roberts, 2020a). This is 8-10% of the lives lost on the estimates of Miles et al. (2020). Medical discoveries are not linearly predictable, of course. Still, we should not make policy while ignoring that they happen. To get an idea of how to model for them, we could for example suppose, using this single data point, that one such treatment might be discovered every 3 months, take the lower (8%) figure, and assume that the treatments' effectiveness is random and independent, so that with M such treatments available the mortality rate is $(0.92)^M$ of the rate at the start of the pandemic. The mortality rates in mid-September 2020, mid-December 2020 and mid-March 2021 would then be reduced by factors of 0.92, 0.84 and 0.78.

Testing is also becoming cheaper, swifter and more widely available; contact-tracing options are improving; knowledge about risky activities and locations is growing; mask-wearing and other ways of reducing transmission may be becoming more prevalent; citizens may be better informed and better able to self-protect. There seems some optimism (which, if rational, implies some chance) that, combined with targeted interventions short of a full lockdown, these factors may – perhaps by mid-September -- allow a relatively low infection rate to be sustained, **once** such a rate is attained.

*2.2 Option value*

The last set of points (4) all have the same form: different states in mid-September may give us different options. The expected value in lives or QALYs saved is uncertain, because this depend on uncertain future developments. But it is irrational to assign them no value. One has to take some view of the probabilities of favourable future developments and the magnitudes of their impacts. Reasonable people can differ about the numbers, but it seems highly implausible that that there is negligible long-run value difference between the states of raging epidemic and near-suppression in mid-September.

I pressed the authors on this point after reading the first draft of (Miles et al., 2020) and thank David Miles for a helpful and collegial discussion. Clearly, though, I was not very persuasive, since their revised draft adds (as far as I can see) only the following comments in response:

"If it was obvious that starting from the first scenario three months ahead (continued lockdown) was obviously better than the scenarios where more people were infected (the three easing scenarios) then it would not be a complete analysis to just focus on costs and benefits over the first three months. But in fact it is not obvious that starting from lockdown three months ahead with low numbers infected is better. The numbers of people who had ever been infected would be lower than under the other scenarios and so the susceptible population would be greater so that the impact of then easing restrictions, in the absence of a vaccine, would be worse."

I agree this is a relevant consideration. Suppose, for the sake of discussion, that we are certain that there will be no vaccine and no future advances in treatment of Covid-19. Suppose too that nothing helpful and new will be learned about its propagation, and there will be no relevant future behavioural changes in the population that reduce the effective R number while keeping society functioning well and – most relevantly – reduce the proportion of the population ( 1-(1/R) ) that need to be infected to effectively reach herd immunity. Under these very pessimistic assumptions, any temporary lockdown only delays the inevitable. Unless society goes into permanent lockdown, then, sooner or later, the infection will propagate through the population until herd immunity is reached. Suppose too that immunity is permanent once acquired. In this (over) simplistic analysis, having more infections "out of the way" is a net gain to be considered in an end state. The lives thus lost are, of course, a net cost, but they have already been counted in the analysis.

However, this is not a very probable set of assumptions. There have already been advances in treatment (Horby et al, 2020) and other treatment options appear promising enough for further exploration (e.g. Rowlett, 2020; Roberts, 2020b). It would be surprising if no further treatment options prove valuable in future. Many research teams are working on a variety of vaccines, with some apparently promising early results. The UK, like other countries, has seen major behavioural changes, which continue to be widespread after lockdown – social distancing, improved hand sanitisation and mask-wearing are all common. Studies are producing better data about the relative effectiveness of different levels of distancing, types of mask, forms of ventilation, air filtering and disinfection, and identifying better which factors cause individuals to be at risk of death or serious illness from Covid-19. All of these developments suggest that delay is likely to be valuable.

Another issue is that, in realistic epidemic models, it is not the case that a fixed fraction of the population becomes infected irrespective of the epidemic trajectory. There is generally an epidemic overshoot, meaning that more than $\left(1 - \left(\frac{1}{R}\right)\right)$ of the population become infected. The expected level of epidemic overshoot needs to be considered in comparing end states, and this requires some assumptions about the future course of the epidemic for each end state.

The key problem with Miles et al.'s comments is that, surprisingly, they ignore the probabilistic nature of option values. Miles et al. seem to be arguing that it is not **certain** that starting from the first scenario in mid-September will turn out, in the long run, to produce fewer deaths than starting from (in particular) their most pessimistic ease scenario. Indeed, it is not **certain.** It is **possible** that there will be no further developments in treatment, no more effective social or behavioural interventions, and no useful vaccine over the course of (say) the next 2-3 years, that we have no option better than allowing the epidemic to propagate through the UK population until we reach herd immunity, and that all the plausible epidemic curves starting from the various mid-September rates ultimately result in the same expected number of deaths (though this last is not a given even if epidemics propagate until herd immunity, because of epidemic overshoots). But to justify Miles et al.'s analysis, one needs to believe that there is no clear reason to prefer any of the scenarios. This requires the pessimistic view that the possibility of no positive developments is very low, or more precisely that either the probabilities or the impacts of favourable developments are low enough that the differences in the expected values of the various end states considered are negligible.

Miles et al., or others, may indeed believe this. There is a spectrum of reasonable opinion on these questions. But it is a controversial view on which many others will have informed and differing opinions. If Miles et al.'s conclusions do indeed rest on these pessimistic beliefs, as they seem to, this needs to be very clearly highlighted and argued.

In principle, one could try to improve the analysis by producing a full stochastic model of future interventions, estimating all the probabilities and impacts of favourable developments. Another way of arriving at a cash figure would be to poll policymakers: if a mercenary genie offered to replace a state of raging epidemic in mid-September by a state of near suppression, how much should we be willing to pay? Both methods are problematic: probability and impact estimates may not be very persuasive, and the alternative still effectively relies on policymakers using something like probability and impact estimates, if perhaps in a not fully consciously articulated mental model. But either seems preferable to simply ignoring the issue.

*2.3 Arbitrarily fixing time frames leads to inconsistent results*

Another way of seeing that there is something very wrong with Miles et al.'s methodology is to ask why they consider a fixed 13-week period, and whether the conclusions would be affected if different periods were considered. Evidently, there is nothing magical about 13 weeks: lockdowns can be shorter, or longer. Indeed, cycles of lockdown/release with periods much shorter than 13 weeks have been proposed as serious policy options (e.g. (Chowdhury et al, 2020)). The UK had many options in mid-June beyond a further 13-week lockdown or a 13-week release.

Miles et al. do not say, but presumably believed that, whatever policy were adopted for the period from mid-June to mid-September, essentially the same analysis could be applied again in mid-September to decide policy for the following 13 weeks. Presumably they also believed that the same conclusions would necessarily follow, and that their fixed period analysis implies that lockdowns will always be more costly than the benefits warrant, although again this is not discussed in (Miles et al., 2020). In fact, even this is not immediately clear, even if we restrict to the binary choice each quarter of lockdown or release. In mid-September, Miles et al.'s third ease scenario produces 7572 weekly deaths, increasing exponentially by a factor of 1.15 per week, implying about 300000 deaths over the following quarter. A lockdown from mid-September in their model would reduce the weekly deaths exponentially by a factor of 0.7 per week, implying about 17500 deaths, so that easing from the quarter following mid-September implies about 282500 excess deaths. If we take Miles et al.'s lowest estimate for the March-June lockdown costing £200bn, 10 QALY per death, and value a QALY at the NICE figure of £30,000, then the QALY costs are £84bn. So, indeed, this calculus favours continued easing, but not by a large factor for this conservative case. As Miles et al. discuss, there is a reasonable case for trebling the QALY value, and this would alter the conclusion.

More importantly, though, if we consider repeated lockdown decisions at fixed intervals, there is no guarantee that the model and methodology will produce consistent answers. This is easy to see in a toy model. Let us assume that the cost of lockdown for N weeks is N times the cost for one week, which we call L. Let us call the cost of a death C. Suppose that we start with M deaths in week zero. Suppose that lockdown reduces the deaths, multiplying by a factor $F_0 < 1$ per week and easing increases them, multiplying by a factor $F_1 > 1$ per week Consider two strategies:

(i) deciding week by week whether to lockdown for the following week,

(ii) deciding at the start whether or not to lockdown for the following N weeks.

Suppose that strategy (i) supports easing in week 1 (as it would with Miles et al.'s parameters). In week 2, it supports continued easing so long as

$$CM \, F_1 \, (F_1 - F_0) < L. \quad (1)$$

If it supports continued easing up to week (N-1), then in week N, it supports continued easing so long as

$$CM \, (F_1)^N \, (F_1 - F_0) < L \, . \quad (2)$$

Strategy (ii) supports easing for all N weeks so long as

$$C\,M \left( F_1 \frac{F_1^N - 1}{F_1 - 1} - F_0 \frac{F_0^N - 1}{F_0 - 1} \right) < N\,L \,. \quad (3)$$

If Equation (2) holds in week N then it holds in all previous weeks (and so implies equation (1)).

It is easy to find parameters such that equation (2) tells us that a weekly decision tree will produce the outcome that we should ease every week during the N weeks, while equation (3) implies that, if we make a one-off decision at the start of the N weeks, we should lockdown throughout.

For a simple example, take $L = CM$, $F_1 > 1$ and $N, F_0$ such that $F_1^N \, (F_1 - F_0) < 1$.

This makes no sense. We cannot trust a methodology whose policy advice depends on an arbitrary choice of time frame.

3. **Illness matters: the QALY costs of non-mortal COVID-19**

Another surprising lacuna in Miles et al.'s analysis is that, while they consider the loss of QALYs arising from COVID-19 deaths, they ignore the loss of quality of life caused by the illness itself and by the after-effects. They also do not consider the possible loss of life expectancy for those who suffer serious COVID-19 illness and survive.    These costs are presently highly uncertain, but this is also true of the other costs and benefits that Miles et al. consider.    And many more people become seriously ill with COVID-19 than die of it, so it is certainly not immediately obvious that the illness costs are negligible compared to, or even smaller than, the mortality costs.

According to the ZOE covid symptom study, as reported by Greenhalgh, Knight, A'Court , Buxton &  Husain (2020) , most people recover from COVID-19 within two weeks, one in ten people may still have symptoms after three weeks, and some may suffer for months.    Other estimates of the proportion of long-term sufferers are higher (Tenforde et al., 2020).

One estimate (Prosser et al., 2006) assigns a cost of 0.005 QALYs for a bout of influenza.    The course of COVID-19 illness appears to be very variable.    Impressionistically, one might guess that it is on average significantly worse than flu even for those not hospitalized, and on average lasts longer than flu.    Giving a factor of 2 for each of these would give a cost of 0.02 QALYs for an average COVID-19 bout.    At a mortality rate of 0.6%, there are about 150 bouts of COVID-19 (costing a total of 3 QALYs) for each death (for which Miles et al. assign a cost of 5-10 QALYs).    Although the illness cost thus estimated is smaller than the mortality costs, it is comparable, already suggesting that illness costs cannot be neglected in a quantitative analysis.

Another approach might be to focus on hospitalization.    About 125000 COVID-19 patients were hospitalized in the UK up to mid-June, while there were about 40000 deaths (not all of which occurred in hospitals). Although an illness requiring hospitalization presumably significantly decreases quality of life during hospitalization, and some COVID-19 survivors are hospitalized for lengthy periods, the average length of hospitalization is short: Rees et al. (2020) suggest ~5 days outside China from early data.    Even assigning a quality of life of 0 during hospitalization, this gives a loss of only ~2000 QALYs from hospitalizations, compared to 200000-400000 from deaths.    Hospitalization QALY costs, per se, thus do seem to be relatively negligible.

The QALY costs of COVID-19 after-effects are hard to estimate.    Lung scarring and continuing breathlessness are reported to be common, even among cases where the symptoms experienced were not severe, with accompanying risk of long-term impairment of lung function (George, Wells & Jenkins, 2020; Shi et al., 2020; Pan et al., 2020).

Long-term symptoms include "cough, low grade fever, and fatigue, … shortness of breath, chest pain, headaches, neurocognitive difficulties, muscle pains and weakness, gastrointestinal upset, rashes, metabolic disruption (such as poor control of diabetes), thromboembolic conditions, and depression and other mental health conditions" (Greenhalgh et al., 2020). Cardiopulmonary complications, thromboembolisms, ventricular dysfunction and neurological sequelae are also reported (Greenhalgh et al., 2020). The long-term prognosis for COVID-19 patients with these various after-effects, which are often combined, is not known.

Producing even a ballpark estimate of these costs is thus very hard, but it is not hard to see that they might be significant. One might expect that there are more patients with severe after-effects than there are deaths. Say that 2% of COVID-19 patients had severe after-effects, diminishing their life quality by 0.2 for 5 years and their life expectancy by 2 years. This would give (2/0.6) x 3 = 10 QALYs from this cohort for each death (for which, recall, Miles et al. assign a cost of 5-10 QALYs).

All the figures in this section are crude guesses. Taken at face value, they would suggest that the QALY cost of COVID-19 is double or more that assigned by Miles et al. However, the numbers used here rely on thin data and guesswork. My only clear conclusion is that, absent more detailed examination of the illness costs, Miles et al.'s quantitative estimates of QALY costs cannot be trusted, and nor can their conclusions. Better data and better analyses of these questions are needed.

4. **Conclusions**

It is very difficult to estimate the economic costs of lockdowns to date or to compare these against the benefits they have produced in avoiding COVID-19 deaths and illnesses. It is also very difficult to estimate the costs lockdowns have produced in deaths and illnesses of other types. There may also have been health benefits of lockdowns: these too are very difficult to estimate. It is even harder to estimate the costs and benefits of future lockdown or easing policies. The medical data about death and illness rates of COVID-19 are still uncertain, and the future prognoses for the more severely affected COVID-19 patients are very uncertain. Any estimates are thus necessarily going to be rough and uncertain. Comparisons of money and life are controversial, and there are reasonable arguments for assigning various values to a QALY.

Miles et al. set out a framework which may be a useful starting point for such discussions, carefully acknowledging several, though not all, of these uncertainties. The quantitative estimates they present lead them to suggest that, even with a value of QALY much higher than the NICE figure, the March-June lockdown in the UK was not, with hindsight, justifiable. They suggest even more strongly that a continued lockdown cannot be justified, even if the alternative (which might plausibly be a worst-case scenario) is an epidemic exponentially growing from mid-June onwards. They suggest that there is enough margin for error in their calculations that there seems little room to doubt their conclusions.

As a starting point for academic discussion, giving a framework for others to improve, this is all valuable. As policy advice, or a contribution to public debate, in its present form, it is highly irresponsible. Their work neglects two key factors, each of which likely makes a large difference to their numbers, and which combine in a way that makes a further large difference, with all these effects adding to the case for lockdowns. It is not at all clear that their conclusions would still be same if these factors were properly incorporated.

One cannot usefully make lockdown policy by ignoring everything that will happen beyond a 3-month horizon, unless one knows the world is going to be hit by an asteroid at that point. (And if it is, the calculations and arguments would be very different.)

Nor can one sensibly ignore the costs of COVID-19 illness and after-effects. These may possibly be larger than the costs of COVID-19 deaths.

This is not to say that either the original lockdown or continuing it beyond mid-June would necessarily have been justified, within the framework considered here, given a model that appropriately incorporates these factors. It is beyond my scope to try to answer these questions here. Building such a model needs informed insights into plausible future developments in the treatment and prevention of COVID-19 and into the long-term health effects of COVID-19. I hope the work of Miles et al. may be extended to address these points and give more trustworthy (albeit very likely uncertain and conditional) answers.

While our focus here has been on one analysis applied to the case of the UK, the logical points apply quite generally.   Rational public policy in tackling COVID-19 cannot be made without estimates for the probabilities of future medical interventions and of long-term illness costs.   More work on modelling these uncertainties is urgently needed.   Any framework for policy analysis should include explicit statements about the underlying assumptions regarding these uncertainties.   In particular, while it might be defensible to make policy recommendations based on the assumption that there is negligible probability of an effective vaccine or other effective new treatment in the foreseeable future, such an assumption needs to be very clearly highlighted, to avoid potentially misleading policy makers and others who may reasonably have a different view of the probabilities.

In principle, these points apply to public policy in tackling any disease.    Indeed, once diseases are well understood, the long-term illness costs are generally reasonably reliably estimated and factored into careful policy analyses.   The option values of uncertain medical developments may often be neglected in because, once medical research into them is in a mature phase, the rate of development of innovative treatments tends to be relatively slow.   However, illness costs and option values should always at least be considered, even if only to argue that they are almost certainly relatively negligible.    When the uncertainties are great, as when tackling novel pathogens such as COVID-19, such arguments are unlikely to be justifiable, and probabilistic analyses are crucial.

**Acknowledgements**    I thank David Miles for a helpful and informative correspondence and anonymous referees for helpful suggestions.